\documentclass[aps,prb,twocolumn,superscriptaddress]{revtex4}
\usepackage[latin1]{inputenc}
\usepackage[english]{babel}
\usepackage{setspace}  
\usepackage{graphicx}
\usepackage{amssymb}
\usepackage{amscd}
\usepackage{rotating}
\usepackage{color}
\def\gapx{\lower 2pt \hbox{$\buildrel>\over{\scriptstyle{\sim}}$\ }}
\def\lapx{\lower 2pt \hbox{$\buildrel<\over{\scriptstyle{\sim}}$\ }}
\begin{document}

\preprint{}

\title{Disorder-induced superfluidity}

\author{Long Dang }
\affiliation{Department of Physics, University of Alberta, Edmonton, Alberta, Canada, T6G 2J1}
\author{Massimo Boninsegni }
\affiliation{Department of Physics, University of Alberta, Edmonton, Alberta, Canada, T6G 2J1}
\author{Lode Pollet}
\affiliation{Theoretische Physik, ETH Z\"urich, 8093 Z$\ddot{u}$rich, Switzerland}

\date{\today}

\begin{abstract}
We use quantum Monte Carlo simulations to study the phase diagram of hard-core bosons with short-ranged {\it attractive} interactions, in the presence of uniform diagonal disorder. It is shown that moderate disorder stabilizes  a glassy  superfluid phase in a range of values of the attractive interaction for which the system is a Mott insulator, in the absence of disorder.  
A transition to an insulating Bose glass phase occurs as the strength of the disorder or interactions increases. 
\end{abstract}

\maketitle
\section{Introduction}
The interplay of superfluidity (SF) and localization in disordered Bose systems has been the subject of intense study for two decades \cite{fisher89}. Most of the theoretical investigative effort has been focused on the quantum phase transition between a superfluid and insulating phase. For example, it has been established that disorder leads to the appearance of an (insulating) Bose glass, sandwiched between the superfluid and Mott insulating phases. This topic is enjoying continued interest, especially since cold atom physicists have recently produced controllable disorder using laser speckles~\cite{ Sanchez-Palencia, White08} and looked at such phenomena as Anderson localization of a one-dimensional condensate~\cite{Roati, Sanchez-Palencia} and the suppression of the condensate fraction in three dimensions.\cite{White08} 

The idea of ``superglass" has come to the forefront in the context of the investigation of the (super)solid ${}^4$He. The superglass phase was initially observed in quantum Monte Carlo simulations, in which the superfluid phase had an inhomogeneous condensate map on a microscopic scale.\cite{superglass} Biroli {\it et al.} proved that such a superglass phase does exist (at least as a metastable phase) by introducing an (artificial) model which could be  mapped to a classical system of hard spheres and studied in a controlled fashion.\cite{biroli} Recent experiments on solid $^4$He have confirmed the strong interplay between a superfluid component and a slow (glassy) dissipative component.\cite{Seamus} However, little is known yet about the superglass phase, and specifically about the actual role of disorder in promoting or enhancing superfluidity. Give the current controversies and puzzles surrounding the interpretation of experiments on the possible supersolid phase of helium, further investigations of superfluid glassy phases are thus warranted.

In this work, we provide strong numerical evidence for disorder-induced superfluidity in a lattice realization of  hard-core bosons with a strong nearest-neighbor attraction, in the presence of external disorder.  In particular, we show that 
at low temperature and in a small range of attractive interactions, disorder of sufficient strength stabilizes a ``glassy" superfluid phase. The superfluid density reaches a maximum and then decays as the strength of the disorder increases, as an insulating glassy phase intervenes. 
In contrast to the case of repulsive bosons where disorder reduces the size of the superfluid phase, we see that strongly attracting hard-core bosons can be stabilized and made superfluid by disorder. 
In other words, disorder {\it induces} superflow in an otherwise insulating phase.  Aside from supersolid $^4$He,  such a scenario  is possibly relevant to  other condensed matter systems, e.g., high-temperature superconductors,\cite{diso} as well as to the elusive superfluid phase of molecular hydrogen,\cite{joe} and to the role of substrate disorder in the superfluidity of (sub)monolayer helium films.\cite{chan03}

\section{Model}
We describe a disordered Bose system by means of  the following Hamiltonian:
\begin{equation}
H= -t\sum_{\langle ij\rangle}(\hat{a}_i^{\dagger}\hat{a}_j + h.c.) +  V\sum_{\langle ij\rangle}\hat{n}_i\hat{n}_j +
  \sum_i  \delta_i  \hat{n}_i\ . \label{ham}
\end{equation}
We consider here a square lattice of $N=L\times L$ sites, with periodic boundary conditions. 
The sums $\langle ij\rangle$ run over all pairs of nearest-neighboring lattice sites, $\hat a^\dagger_i$ $(\hat a_i)$  is the Bose creation (annihilation) operator for a particle at site $i$, $\hat{n}_i=\hat{a}_i^{\dagger}\hat{a}_i$ is the local density operator.  The first term of (\ref {ham}) represents the hopping of particles to nearest-neighboring sites. Henceforth, we choose the hopping integral $t$ as our energy unit. 

The second term represents the interaction among bosons.  A {\it hard-core} on-site repulsion is assumed, limiting the occupation of every site to no more than one particle. For the nearest-neighbor interaction, essentially all previous work based on (\ref{ham}) has focused on  the {\it repulsive} case, i.e. $V > 0$, chiefly to elucidate the nature of the disorder-driven superfluid to insulator transition.\cite{group} An {\it enhancement} of superfluidity by disorder has been predicted in some cases.\cite{krauth} Here, on the other hand, we consider the case of {\it attractive} nearest-neighbor interaction (i.e., {\it negative} $V=-|V|$), and neglect interactions among particles lying at distances greater than nearest neighbors. With this choice,  the Hamiltonian (\ref{ham}), which is essentially a lattice model of quantum ``sticky" spheres, is isomorphic to that of a  spin-1/2 XXZ quantum ferromagnet.  

We model disorder by means of a random on-site potential $\delta_i$, uniformly distributed in the interval $[-\Delta,\Delta$]. Other theoretical representations of a disordered environment could be considered, e.g., one in which the hopping matrix element $t$ randomly varied from site to site, but in this work we restricted ourselves to the above, widely adopted {\it diagonal} model of disorder.\cite{krauth,group} In the spin language, the disordering potential is equivalent 
to a random on-site magnetic field along the $z$ axis. 


In the absence of disorder, the ground state of (\ref{ham}) is a superfluid for $|V| < 2$, whereas for $|V| \ge 2$ only a Mott insulating phase exists, with exactly one particle per site, regardless of lattice geometry and dimensionality. This is simply because the system can maximally lower its energy by having each particle surrounded by as many nearest-neighboring particles as possible, trumping any contribution from the hopping term. The regime of interest in this work is the latter, i.e., that in which no superfluid phase exists in the absence of disorder.  


\section{Methodology}
We perform grand-canonical quantum Monte Carlo simulations to study the ground state properties of (\ref{ham}), using the Worm Algorithm in the lattice path-integral representation.\cite{prokofiev98,lode07}
As the details of this computational method are extensively described elsewhere, and because the calculations performed here are standards, we shall not review it here, and simply refer interested readers to the original references.

The results shown here correspond to a temperature $T$ sufficiently low (typically $\beta=1/T=L$), to be regarded as essentially ground state estimates.  Simulations are carried out over square lattice of size varying from $L$=12 to $L$=96, and estimates are averaged over a number $M$ of independent realizations of the random disordering potential, typically $M$=100 (20) for $L$=12 (96).

\section{Results}

\begin{figure}
\includegraphics[scale=0.32,angle=0]{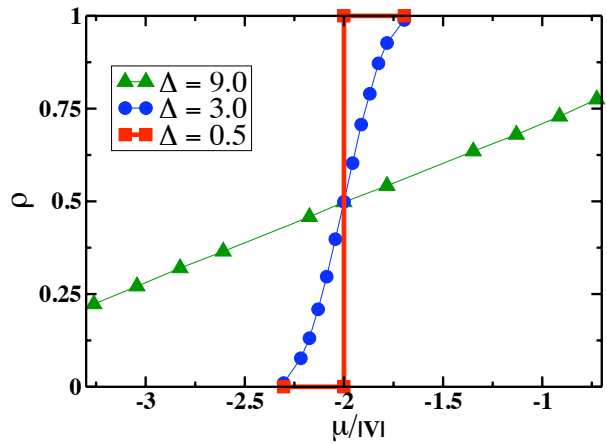}
\caption{(Color online). Ground state density $\rho$ versus chemical potential $\mu$ for $|V|$=2.3, for weak  ($\Delta$=0.5, squares), intermediate ($\Delta$=3.0, circles) and strong ($\Delta$=9.0, triangles) disorder. Results shown are for a lattice of size $L$=96, and are obtained by averaging over 20 independent realizations of the disorder. Statistical errors are smaller than symbol sizes.}\label{f1}
\end{figure}

Fig. \ref{f1} shows the average particle density $\rho$ as a function of the chemical potential $\mu$, for a particular value of  $|V|$ greater than 2 ($|V|$=2.3). For weak disorder (i.e., small $\Delta$), the ground state of the system has exactly one particle per site, with an abrupt density jump at $\mu/|V|=2$, when the lattice turns from empty to fully filled. However, for disorder of sufficient strength (figure shows results for 
$\Delta = 3$), the density jump disappears, being replaced by a smooth curve, signaling continuous dependence of  density on chemical potential. In other words, the disorder stabilizes phases at intermediate densities, consisting of interconnected ``clusters" of particles, pinned by local fluctuations of the  disordering potential. In this situation, the value $\mu/|V|=2$ corresponds to a particle density $\rho=0.5$. 

\begin{figure}
\includegraphics[scale=0.32,angle=0]{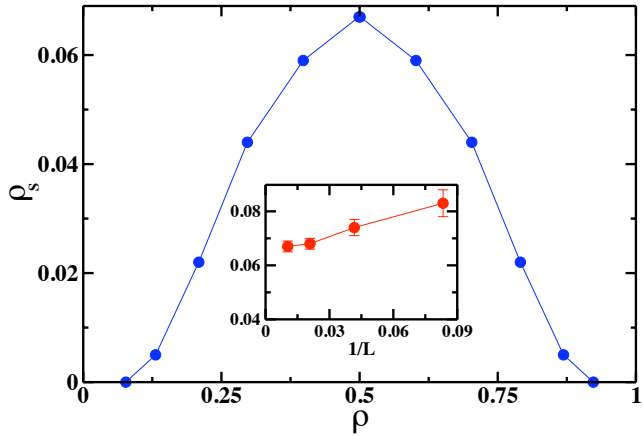}
\caption{(Color online). Superfluid density $\rho_S$ versus particle density $\rho$ for $|V|$=2.3 and disorder strength $\Delta=3$. Statistical errors are smaller than symbol sizes. Results shown are for a square lattice with $L=96$, and $\beta$=$L$, and are obtained by averaging over 20 independent realizations of the disorder. The solid line is a guide to the eye. {\it Inset}: Superfluid density for a fixed particle density $\rho$=0.5, computed on square lattices of varying size $L$. Extrapolation to infinite system size still gives a finite superfluid density. }
\label {f2}
\end{figure}

\begin{figure}
\includegraphics[scale=0.33,angle=0]{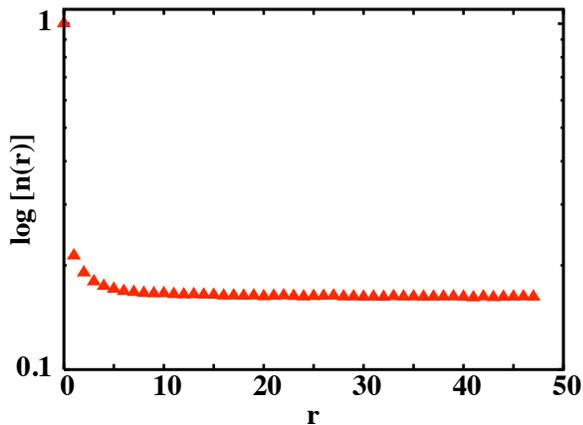}
\caption{(Color online). One-body density matrix $n(r)$  for $|V|$=2.3 and disorder strength $\Delta=3$. Statistical errors are smaller than symbol sizes. Results shown are for a square lattice with $L=96$, and $\beta$=$L$, and are obtained by averaging over 20 independent realizations of the disorder. Data show a weak power law decay of $n(r)$ at long distances.}
\label {f3}
\end{figure}

Clearly, the issue immediately arises of whether such disordered phases may turn superfluid at low $T$, and what the nature would be of such a disordered superfluid phase, 
simultaneously featuring broken translational invariance.
We investigated the occurrence of superfluid behavior  by directly calculating the superfluid density $\rho_S$ (using the standard winding number estimator).  Fig. \ref{f2} shows $\rho_S$ as a function of particle density $\rho$, in the limit $T$$\to$ $0$, for the one of the choices of model parameters of Fig. \ref{f1}, namely $\Delta=3$ and $\vert V\vert =2.3$.  The superfluid density increases from zero and reaches a maximum value at half filling, where approximately 12\% of the system is superfluid.

Obviously,  numerical data such as those shown in Fig. \ref{f2} must be extrapolated to the $L\to\infty$, in order for us to be able to male confidently the statement that superfluidity observed in these systems is not merely a finite-size effect but survives in the thermodynamic limit. The inset of Fig. \ref{f2} shows a typical extrapolation; estimates are shown for the superfluid density obtained for a fixed particle density $\rho=0.5$, on square lattices of different sizes (12, 24, 48 and 96),  for $|V|$=2.3 and $\Delta$=3. It is worth restating that these estimates are obtained by averaging results corresponding to several independent realizations of the disordering potential.  Based on results such as those shown in the inset of Fig. \ref{f2}, we conclude that the superfluid signal remains finite in the thermodynamic limit. In general, we have observed that results obtained on a lattice with $L$=96 offer a close representation of the physics of the thermodynamic limit, at least in the range of parameters discussed here.  

Additional numerical evidence of superfluidity is extracted from the behavior of the one-particle density matrix $n(r)$, shown in Fig. \ref{f3} for the case $|V|=2.3$ and $\Delta=3$. The data are consistent with quasi long-range order, as expected in two dimensions, namely a power law decay of $n(r)$ at long distances. 
 
\begin{figure}
\includegraphics[scale=0.32,angle=0]{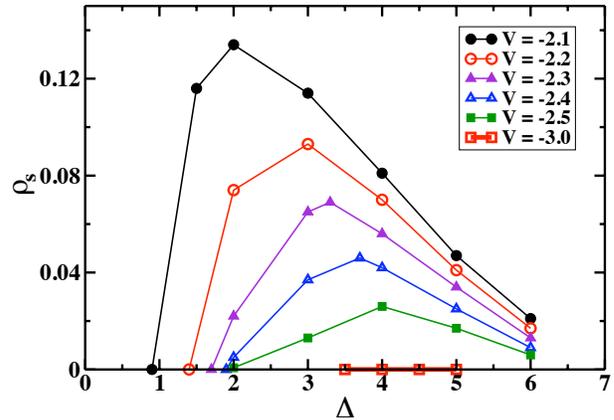}
\caption{(Color online). Maximum value of the  superfluid density $\rho_S$ (attained for $\rho=0.5$) versus disorder strength $\Delta$ for different attractive interactions $V$. Statistical errors are smaller than symbol sizes. Solid lines are only meant to guide the eye.}\label{f4}
\end{figure}

The observed superfluid phase is ostensibly induced by disorder, which stabilizes uniform phases of filling intermediate between zero and one. In order to gain further insight and in-depth understanding of the role of disorder in actually promoting superfluidity, it is of interest to study the competition between the strength of disorder ($\Delta$) and that of the attractive boson interaction ($|V|$). For definiteness, we consider the case  of half filling, corresponding to a maximum in the superfluid density (for those systems for which superfluidity is observed). The same trends are also observed away from half filling.

\begin{figure}
\includegraphics[scale=0.32,angle=0]{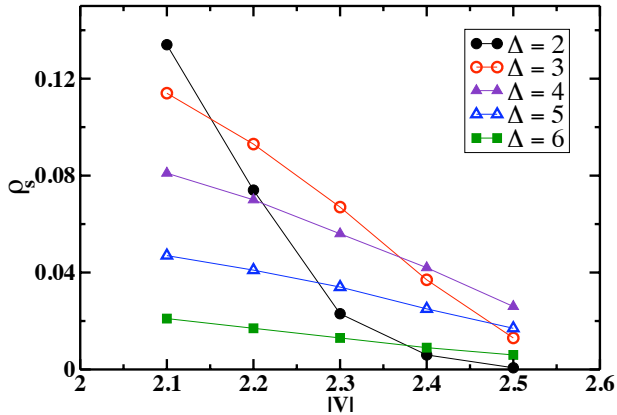}
\caption{(Color online). Maximum value of the superfluid density $\rho_S$( attained for $\rho$=0.5) versus absolute value of attractive interaction $|V|$ for different diagonal disorder $\Delta$ at inverse temperature $\beta=96$. Statistical errors are smaller than symbol sizes. Results shown are for a square lattice with $L=96$. The solid lines are a guide to the eye. }\label{f5}
\end{figure}

When the disorder is weak ($\Delta \ll \vert V \vert $), it cannot break apart clusters of particles, hence  the system remains insulating, as shown in Fig.~\ref{f4} or by the vanishing compressibility $\kappa = d\rho/d\mu$  for $\Delta = 0.5$ in Fig.~\ref{f1}.  There are thus macroscopic domains (empty or fully filled) with hidden long-range order in the system.\cite{Seppala98, Imry75} When the disorder becomes of the order of the attraction ($\Delta \le \vert V\vert $), sites and regions begin to appear throughout the system  where the chemical potential is low enough to rip particles off the cluster, which then breaks down into large grains. These particles, however, are still largely localized in the vicinity of the energetically favorable sites created by disorder, as the curve for $\Delta=2$ in Fig.~\ref{f5} shows.  

 If we further increase the disorder strength,  the grain size decreases to a microscopic scale, a (relatively) large fraction of the particles are delocalized, and superfluidity along interfaces (ridges) becomes possible, as also shown in previous numerical studies.\cite{superglass, Burovski05} This effect  takes place essentially due to percolation. We can thus say that the disorder counters the insulating trend caused by the attractions, and actually makes the system superfluid. This is shown in Fig.~\ref{f4} where we see a rather large superfluid fraction as a function of disorder. Naturally,  as  the disorder strength is increased even further, insulating glassy behavior re-appears, because the disorder is now so strong that it can block any superfluid path and localize particles, much as in the case of repulsive interactions.

A similar scenario takes place on increasing the interaction strength $\vert V \vert$ at constant disorder bound $\Delta$, as shown in Fig.~\ref{f5}. We have already explained the steep decay of the curve corresponding to $\Delta = 2 < \vert V \vert $ above, due to the lack of carriers. When $\Delta$ is greater than $\vert V \vert$, the disorder is sufficiently strong to destroy all macroscopic domains, and superfluidity can occur all over the sample. But as the disorder becomes stronger, it prevents particle world lines from winding around the lattice. 
In the regime of strong disorder, both the disorder and the attractive interactions contribute to suppress superfluidity, as regions with nearly uniform chemical potential will be insulating due to the strong attraction, which pulls particles together in such regions. 

It is worth noting that the above scenario is quite different from that of the repulsive disordered Bose-Hubbard model, where regions of uniform chemical potential are crucial for stabilizing locally a liquid phase, and thus the superfluid properties and the compressibility of the whole system.\cite{fisher89}
We also note here that this insulating phase is compressible,  as shown in Fig.~\ref{f1} for $\Delta = 9t$, which justifies the nomenclature ``Bose glass".\cite{fisher89}  
The compressibility at half filling goes from zero in the phase segregated regime (no disorder), to a large value in the superfluid phase, and decreases then monotonically over the Bose glass phase, when increasing the disorder bound at constant interaction strength.

\begin{figure}
\includegraphics[scale=0.56,angle=0]{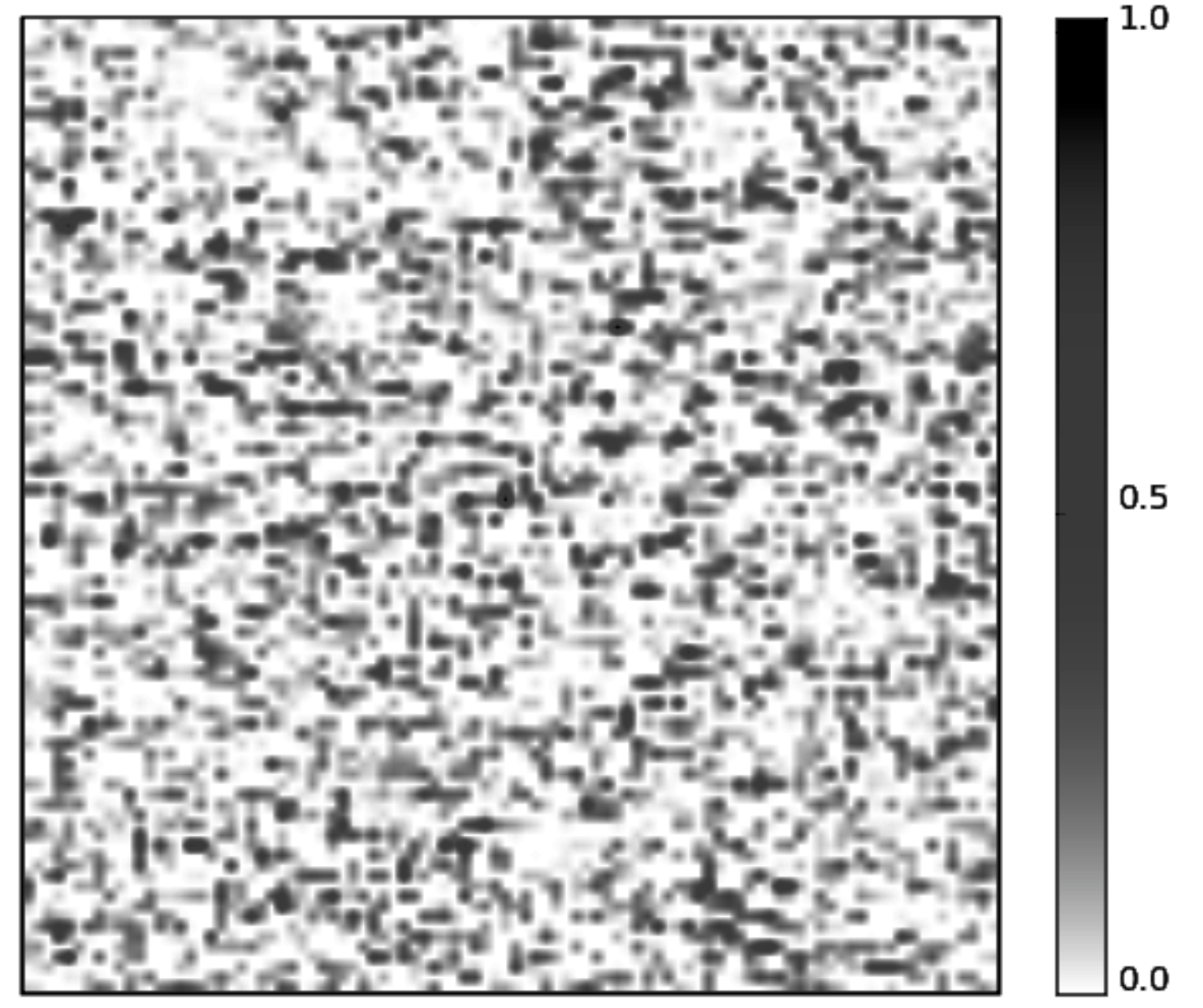}
\caption{Map of the local superfluid density for a particular disorder realization, on a square lattice with $L$=96, $\vert V\vert=2.3$ and $\Delta=3$. The total superfluid density $\rho_S$ equals 0.068(1) for this run. 
The white areas are small insulating grains, connected by superfluid interfaces.}\label{f6}
\end{figure}

This  ``superglass" phase can be visualized through local superfluid density maps, shown in Fig. \ref{f6}, for a particular realization of disorder. The local value is obtained by statistically averaging local contributions to the total superfluid density (i.e., to the square of the winding number), which, in the case shown in Fig. \ref{f6} for $\Delta=3$ and $\vert V\vert=2.3 $, amounts to slightly less than 7\%. We found that the covariance between the superfluid density and the disordering potential is virtually zero, i.e., for these values of the parameters the physics is mostly driven by the attraction between bosons, consistent with the picture given above in the case of strong disorder and strong attraction.

\section{Conclusions}
We have shown that disorder induces a superfluid phase in a lattice system of hard-core bosons with a strong nearest-neighbor attraction. While the system without disorder is an insulator of the ferromagnetic Ising type, the disorder can induce an inhomogeneous superfluid (or superglass) phase (corresponding to in plane order in the spin parlance) in a window of interaction and disorder strengths. For stronger disorder bounds, the disorder and the attractive interactions work together to localize the particles.
The coherence induced by disorder might easily be observable in time-of-flight images for ultracold atoms or molecules. 
 
In the absence of disorder, the physics of our model is reminiscent of that of molecular {\it para}-hydrogen, long speculated to be a potential ``second superfluid", due to the light mass of its constituents (one half of that of helium atoms).  On the other hand, superfluidity is not observed in {\it para}-hydrogen due to the strength of the intermolecular potential, which causes the system to crystalize at temperatures significantly above that at which Bose Condensation is expected to take place.\cite{ginzburg} Recent numerical studies\cite{joe} have shown that disorder ought {\it not} give rise to a superfluid phase of {\it para}-hydrogen. Based on the results obtained in this work, we may argue that {\it para}-hydrogen may be a system too ``deep" into the insulating regime (i.e., the effective value of $\vert V \vert$ is too large) for disorder to stabilize a superfluid phase.

On the other hand, the results obtained here suggest that disorder may be responsible for the observation of superfluidity is helium films at coverages corresponding to less than a full monolayer. It should be noted, though, that helium films are expected to be superfluid at ``negative" pressure, in fact all the way down to the spinodal density,\cite{boninsegni} rendering them essentially different than what is discussed here.

\section*{Acknowledgments} This work was supported in part by  the Natural Science
and Engineering Research Council of Canada under research grant G121210893, by the Alberta Informatics Circle of Research Excellence and by the Swiss National Science Foundation. One of us (MB) gratefully acknowledges the hospitality of the Institute of Theoretical Physics, University of Innsbruck.


\begin{thebibliography}{99}
\bibitem{fisher89}M. P. A. Fisher, P. B. Weichman, G. Grinstein, D. S. Fisher, Phys. Rev. B {\bf 40}, 546 (1989).
\bibitem{Sanchez-Palencia} J. Billy, V. Josse, Z. Zuo, A. Bernard, B. Hambrecht, P. Lugan, D. Cl{\' e}ment, L. Sanchez-Palencia, P. Bouyer, and A. Aspect, Nature {\bf 453}, 891 (2008);  L. Sanchez-Palencia, D. Cl{\' e}ment, P. Lugan, P. Bouyer, and A. Aspect, New J. Phys., {\bf 10}, 045019 (2008).
\bibitem{White08} M. White, M. Pasienski, D. McKay, S. Zhou, D. Ceperley, and B. DeMarco, Phys. Rev. Lett. {\bf 102}, 055301 (2009).
\bibitem{Roati} G. Roati, C. D'Errico, L. Fallani, M. Fattori, C. Fort, M. Zaccanti. G. Modugno, M. Modugno and M. Inguscio, Nature {\bf 453}, 895 (2008).
\bibitem{superglass}
M. Boninsegni, N. Prokof'ev and B. Svistunov, Phys. Rev. Lett. {\bf 96}, 105301 (2006).
\bibitem{biroli} G. Biroli, C. Chamon and F. Zamponi, Phys. Rev. B {\bf 78}, 224306 (2008).
\bibitem{Seamus}  J. C.  Davis, private communication (2008).

\bibitem{diso}
See, for instance, J. C. Phillips, PNAS {\bf 105}, 9917 (2008).
\bibitem{joe} J. Turnbull and M. Boninsegni, Phys. Rev. B {\bf 78}, 144509 (2008).
\bibitem{chan03} H. Cho and G. A. Williams, Phys. Rev. Lett. {\bf 75}, 1562 (1995); G. A. Cs\'athy, J. D. Reppy, and M. H. W. Chan, Phys. Rev. Lett {\bf 91}, 235301 (2003).
\bibitem{krauth} W. Krauth, N. Trivedi and D. M. Ceperley, Phys. Rev. Lett. {\bf 67}, 2307 (1991). 
\bibitem{group}
M. Makivic, N. Trivedi and S. Ullah, Phys. Rev. Lett. {\bf 71}, 2307 (1993);
M. Wallin, E. S. Sorensen, S. M. Girvin and A. P. Young, Phys. Rev. B {\bf 49}, 12115 (1994);           
F. Alet and E. S. Sorensen, Phys. Rev. E {\bf 67}, 015701 (2003); 
N. Prokof'ev and B. Svistunov, Phys. Rev. Lett. {\bf 92}, 015703 (2004).
\bibitem{prokofiev98} N. V. Prokof'ev, B. V. Svistunov, and I. S. Tupitsyn, Phys. Lett. A {\bf 238}, 253 (1998); Sov. Phys. JETP {\bf 87}, 310 (1998). 
\bibitem{lode07} K. V. Houcke, L. Pollet and S. M. A. Rombouts, J. Comp. Phys. {\bf 225}, 2249 (2007).
\bibitem{Seppala98} E. T. Sepp\"al\"a, V. Pet\"aj\"a , and  M. J. Alava, Phys. Rev. E {\bf 58}, R5217 (1998).
\bibitem{Imry75} Y. Imry and S. Ma, Phys. Rev. Lett. {\bf 35}, 1399 (1975). 
\bibitem{Burovski05} E. Burovski, E. Kozik, A. Kuklov, N. Prokof'ev, and B. Svistunov, Phys. Rev. Lett. {\bf 94}, 165301 (2005).
\bibitem{ginzburg} V. L. Ginzburg and A. A. Sobyanin, { JETP Lett.} {\bf 15}, 242 (1972).
\bibitem{boninsegni} M. Boninsegni, M. W. Cole and F. Toigo, Phys. Rev. Lett. {\bf 83}, 2002 (1999).





\end{thebibliography}
\end{document}